\documentclass[epj]{webofc}
\usepackage[varg]{txfonts}   

 \newcommand\ra{\rangle}
 \newcommand\beq{\begin{equation}}
 
 \newcommand\eeq{\end{equation}}
 \newcommand\beqn{\begin{eqnarray}}
 \newcommand\eeqn{\end{eqnarray}}
 \newcommand\GeV{\,{\rm GeV}}

\woctitle{XLV International Symposium on Multiparticle Dynamics}
\begin{document}
\title{Breakdown of QCD factorization
in hard diffraction}
%
%

\author{B.Z. Kopeliovich\inst{1}\fnsep\thanks{\email{boris.kopeliovich@usm.cl}} 
}

\institute{Departamento de F\'{\i}sica,
Universidad T\'ecnica Federico Santa Mar\'{\i}a; and\\
Centro Cient\'ifico-Tecnol\'ogico de Valpara\'iso;
Casilla 110-V, Valpara\'iso, Chile
          }

\abstract{Factorization of short- and long-distance interactions is severely broken in hard diffractive hadronic collisions. Interaction with the spectator partons leads to an interplay between soft and hard scales, which results in a leading twist behavior of the cross section, on the contrary to the higher twist predicted by factorization. This feature is explicitly demonstrated for diffractive radiation of abelian (Drell-Yan, gauge bosons, Higgs) and non-abelian (heavy flavors) particles.
}
\maketitle
\section{QCD factorization in diffraction}
\label{intro}

QCD factorization in inclusive processes is nowadays one of the most powerful and frequently used theoretical tools \cite{factorization}. In spite of lack of understanding of 
the soft interaction dynamics, the contributions of the soft long-distance and hard short-distance interactions factorise. Making a plausible (not proven) assumption about universality of the former, one can measure it with electro-weak hard probes (DIS, Drell-Yan process) and apply to hard hadronic processes. Although it is tempting to extend this factorization scheme to diffractive, large rapidity gap processes, it turns out to be heavily broken \cite{rev1,rev2}, as is demonstrated below.

{\sl Ingelman-Schlein picture of diffraction}~\cite{is}.

\begin{figure*}[h]
\centering
  \includegraphics[width=2.2cm,clip]{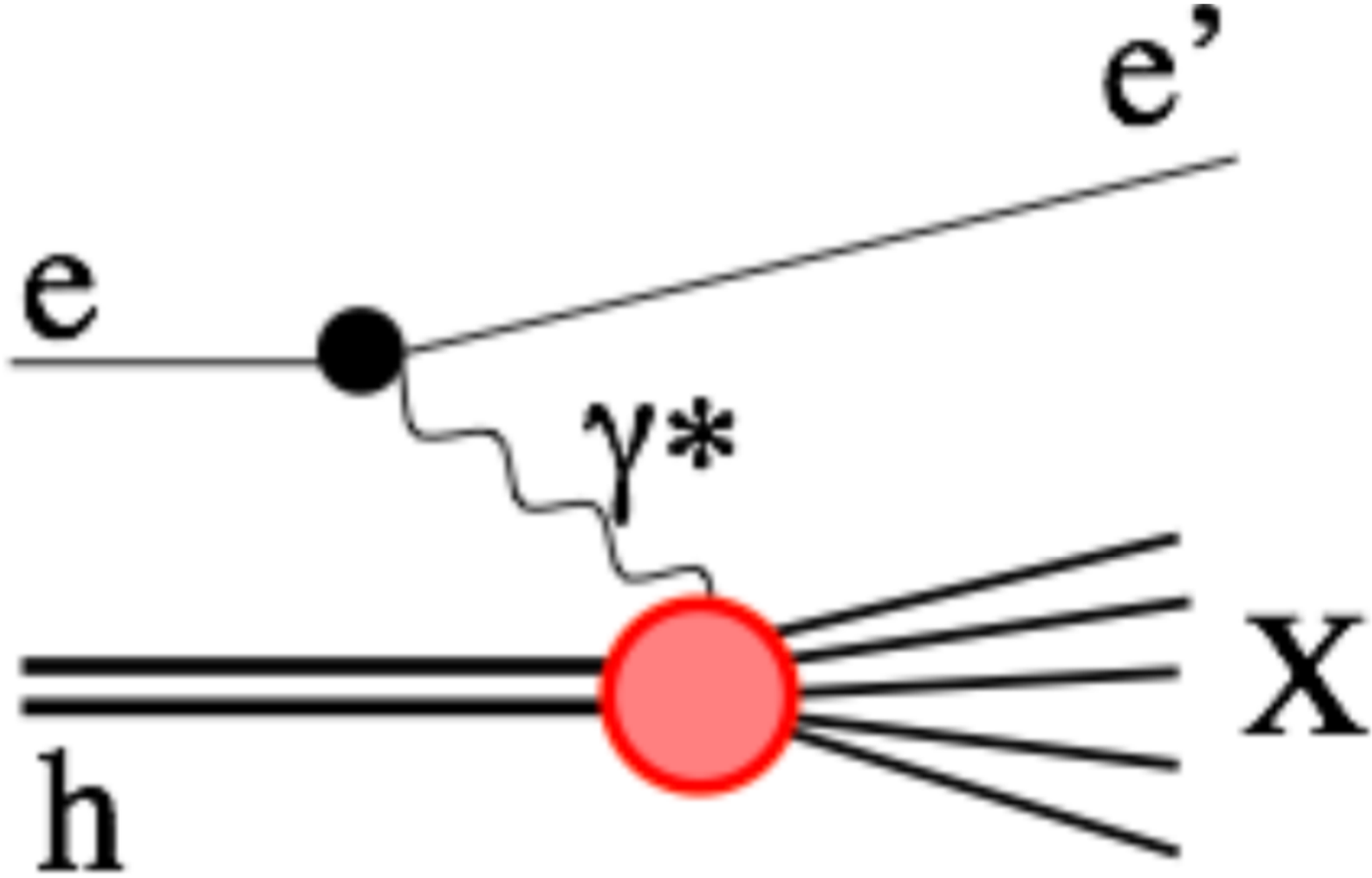}
  \hspace*{2cm}
\includegraphics[width=3cm,clip]{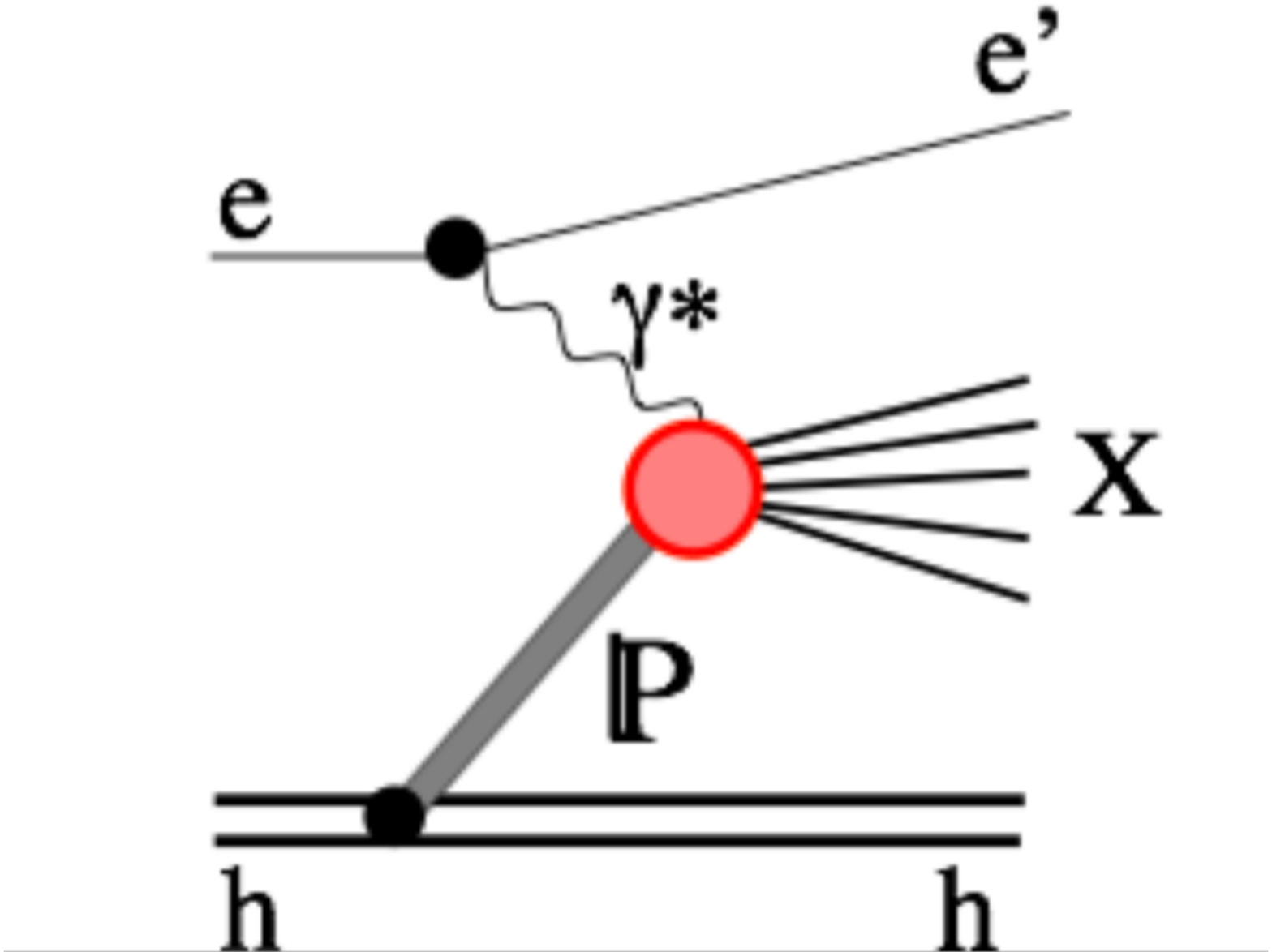}
\caption{DIS on a hadron taget (left) and on the Pomeron, treated as a target (right).}
\label{fig:dis}       
\end{figure*}

It looks natural that on analogy of DIS on a hadronic target, DIS on the Pomeron probes its PDF (parton distribution function), like is illustrated in Fig.~\ref{fig:dis} . 
Once the parton densities in the Pomeron were known, one could 
predict any hard diffractive hadronic reaction. 

{\sl The Good-Walker mechanism of diffraction}~\cite{glauber,FP56,GW}.

According to this quantum mechanical treatment of diffraction, the diffractive amplitude is given 
by the difference between the elastic amplitudes of different Fock
components in the projectile particle. In the dipole representation 
hard diffraction of a hadron comes from the difference between elastic 
amplitudes of hadronic states with and without a hard fluctuation,
\beq
A_{diff}^{h}\propto \sigma_{\bar qq}(R+r)-\sigma_{\bar qq}(R) \propto rR\sim 1/Q,
\label{40}
\eeq
where $\sigma_{\bar qq}(R)$ is the total dipole-nucleon cross section \cite{zkl}; $R$ characterises the hadronic size, while small $r\sim 1/Q\ll R$ is related to the hard process \cite{dy-k,dy-r}. Apparently such a mild $Q$-dependence contradicts factorization prediction, based on the DIS relation,
\beq
A_{diff}\propto \sigma_{\bar qq}(r) \propto r^2\sim 1/Q^2,
\label{20}
\eeq
 which is a higher twist effect.

\section{Drell-Yan reaction: annihilation or bremsstrahlung?}
\label{dy}

Parton model is not Lorentz invariant, interpretation of hard reactions varies with reference frame. E.g. DIS is treated as a probe for the proton structure in the 
Bjorken frame, but looks differently in the target rest frame, as interaction of hadronic components of the photon. Only observables are Lorentz invariant.

\begin{figure*}[h]
\centering
 \includegraphics[width=7cm,clip]{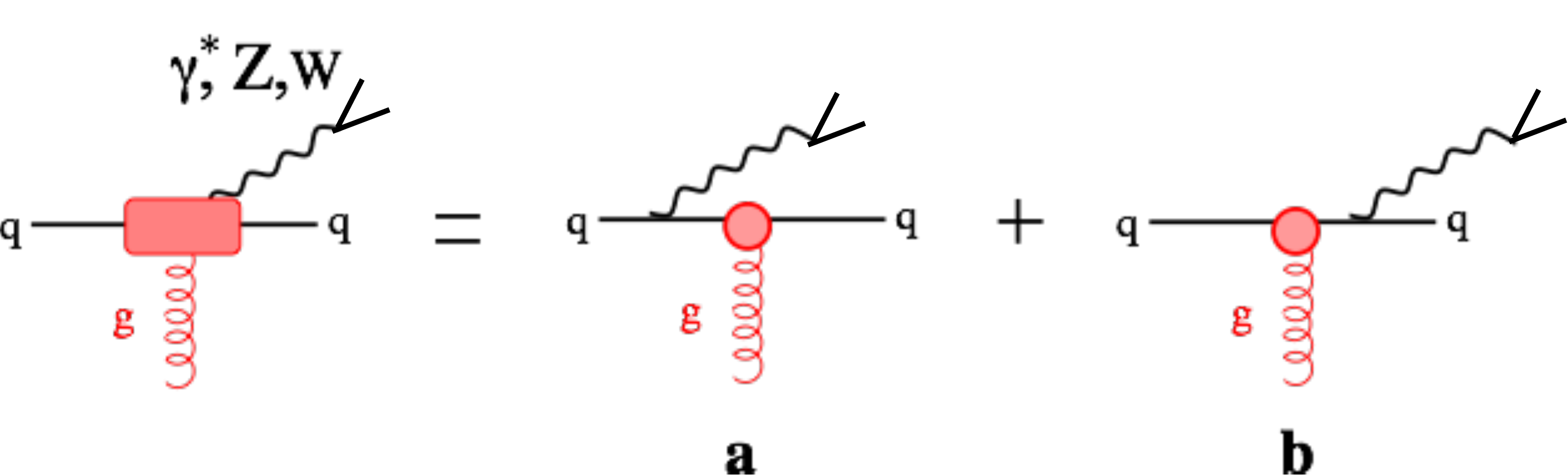}
\caption{Radiation of a heavy photon, or gauge bosons in the target rest frame,  corresponds to $\bar qq$ annihilation in the boson rest frame.}
\label{fig:dy-incl}       
\end{figure*}

The Drell-Yan reaction in the target rest frame
looks like radiation of a heavy photon (or Z, W), rather than q-qbar annihilation
\cite{hirr,kst1}, as is illustrated in Fig.~\ref{fig:dy-incl} 

The cross section, expressed via the dipoles \cite{hirr,kst1}, looks similar to DIS,
\beq
\frac{d\sigma_{inc}^{DY}(qp\to\gamma^*X)}  
{d\alpha\,dM^2} =
\int d^2r\,\left|\Psi_{q\gamma^*}(\vec r,\alpha)\right|^2\,
\sigma\left(\alpha r,x_2\right),
\label{60}
\eeq
where $\Psi_{q\gamma^*}(\vec r,\alpha)$ is the distribution function for the $|\gamma^*q\ra$ Fock component of the quark; $\alpha=p^+_{\gamma^*}/p^+_q$ is the fractional ligh-cone momentum of the heavy photon.

\begin{figure*}[h]
\centering
 \includegraphics[width=9cm,clip]{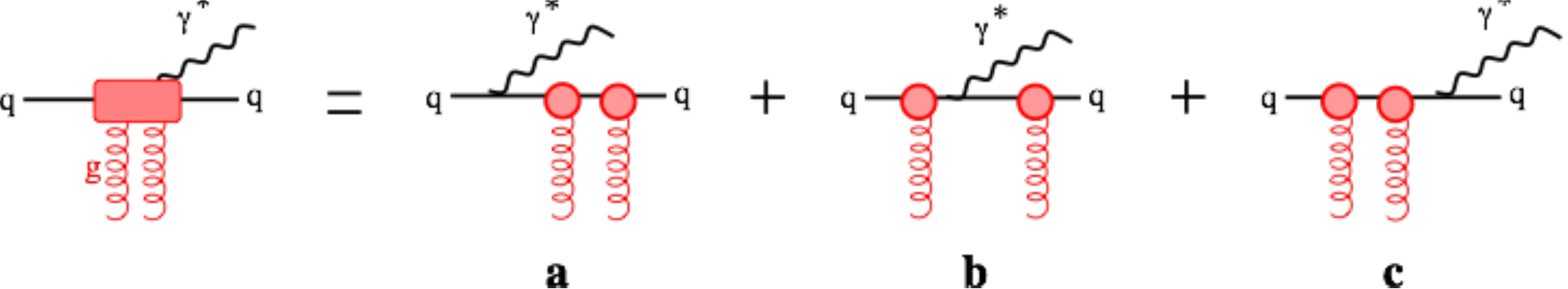}
\caption{Feynman graphs for diffractive radiation of a heavy photon by a quark.}
\label{fig:dy-diff}       
\end{figure*}

In DY diffraction the Ingelman-Schlein factorization is broken.
Indeed, diffractive radiation of an abelian particle vanishes in the forward direction~\cite{kst1}, due to cancellation of the graphs {a}, {b} and {c} depicted in Fig.~\ref{fig:dy-diff},
\beq
\left.\frac{d\sigma_{inc}^{DY}(qp\to\gamma^*qp)}  
{d\alpha\,dM^2\,d^2p_T}\right|_{p_T=0} = 0.
\label{80}
\eeq
In both Fock components of the quark, $|q\ra$ and $|q\gamma^*\ra$ only quark interacts, 
so they interact equally, and according to the Good-Walker picture cancel in the forward diffractive amplitude.
This conclusion holds for any abelian diffractive radiation of $\gamma^*$, $W$, $Z$ bosons, Higgs.

Diffractive DIS is dominated by soft interactions \cite{k-povh, rev1}. On the contrary, diffractive
Drell-Yan gets the main contribution from the interplay of soft and hard scales
\cite{dy-k,dy-r} (see Eq.~(\ref{20})).

\begin{figure*}[h]
\vspace*{-2cm}
\centering
 \includegraphics[width=7cm,clip]{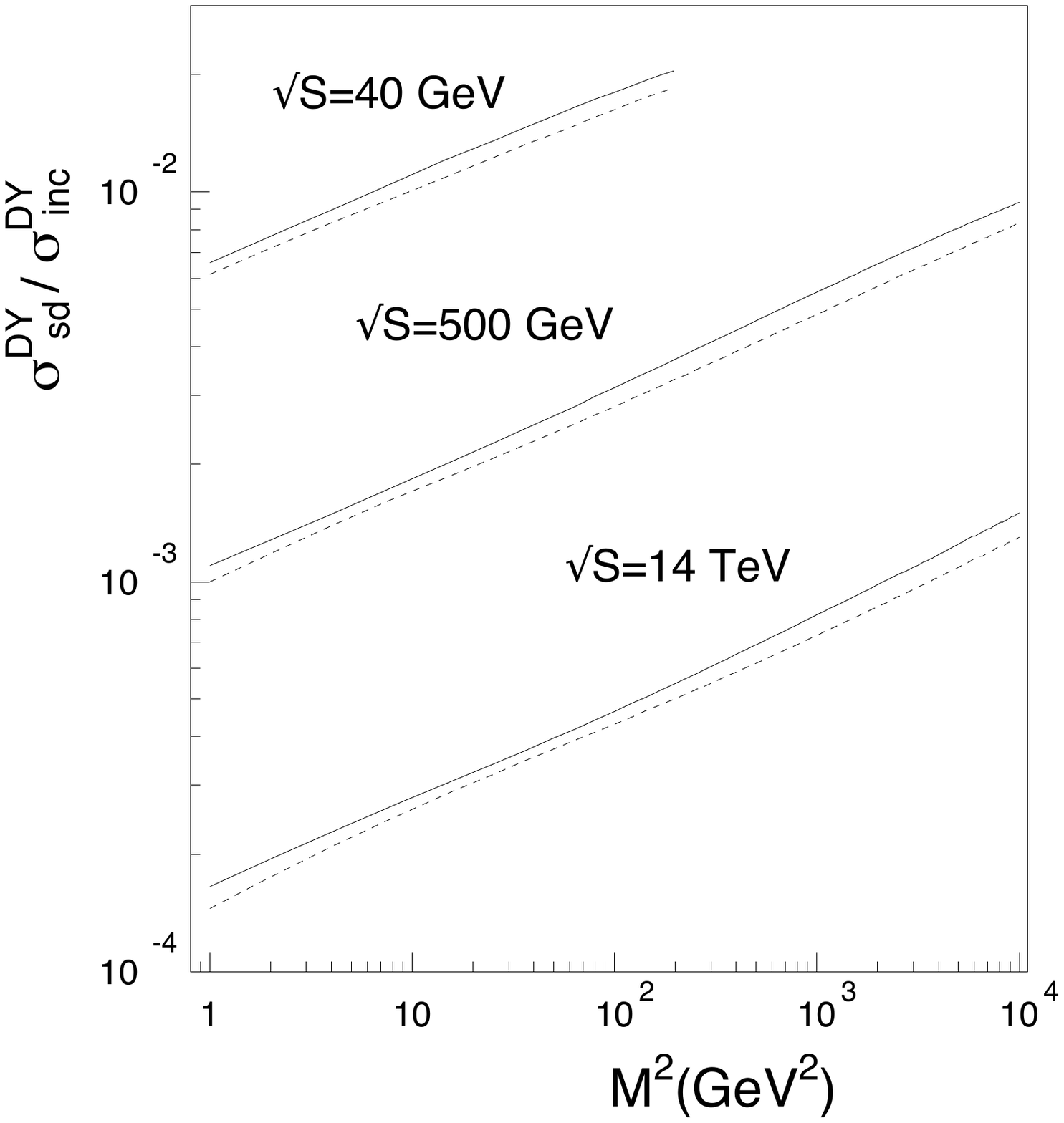}
 \hspace*{-1cm}
\includegraphics[width=7.5cm,clip]{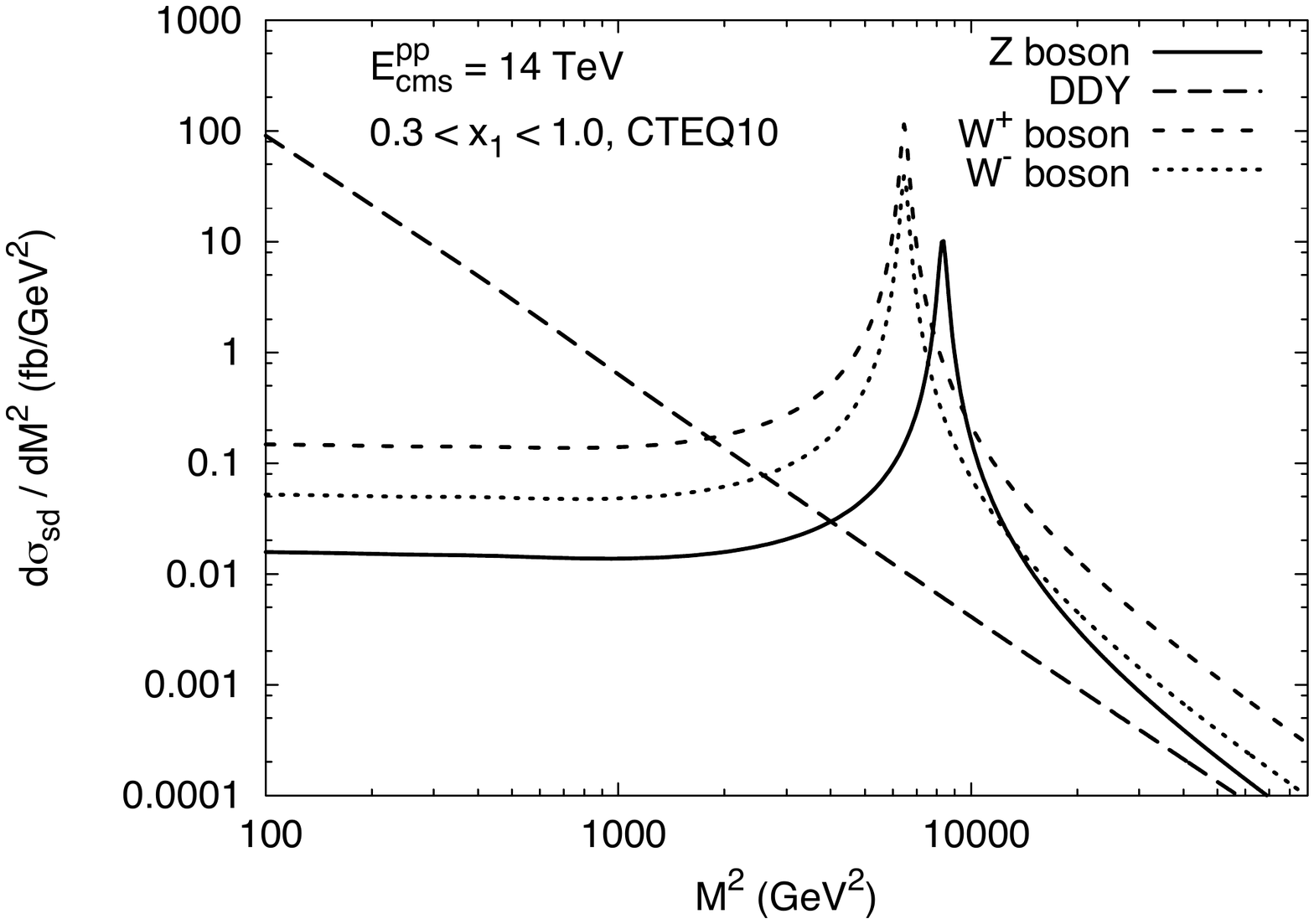}
\vspace*{-1.8cm}       
\caption{{\sl Left:} Fractional DY cross section vs the dilepton mass squared and at different c.m. energies.
{\sl Right:} Diffractive gauge boson production cross section as function of di-lepton invariant mass squared. }
\label{fig:dy-results}       
\end{figure*}

The saturated shape of the dipole 
cross section,  $\sigma(R)\propto 1-\exp(-R^2/R_0^2)$,  leads 
to the unusual features of diffractive Drell-Yan cross section (compare with (\ref{20})), 
\beq
\frac{\sigma_{sd}^{DY}}{\sigma_{incl}^{DY}}\propto
\left[\sigma(R+r)-\sigma(R)\right]^2\propto
\frac{\exp(-2R^2/R_0^2)}{R_0^2}
\label{100}
\eeq
As a result, the fractional diffractive Drell-Yan cross section is steeply falling with energy, but rises with the scale, because of saturation, as is shown in Fig.~\ref{fig:dy-results}.
\vspace{-1cm}

\section{Diffractive Z and W production}
\label{ZW}

Abelian diffractive radiation of any particle is described by the same Feynman graphs, only couplings and spin structure are different
\cite{WZ}. In Fig.~\ref{fig:dy-results}
(right panel) we present the results for
the single diffractive cross sections for $Z^0,\,\gamma^*$
(diffractive DY) and $W^{\pm}$ bosons production, differential in
the di-lepton mass squared, $d\sigma_{sd}/dM^2$.

\begin{figure*}[h]
\centering
 \includegraphics[width=10cm,clip]{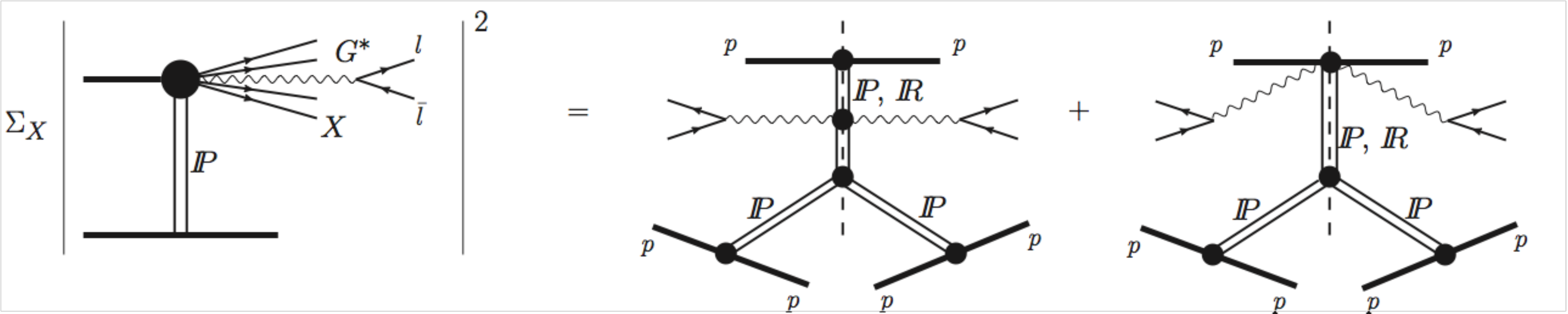}
\caption{Triple-Regge description of the process $pp\to Xp$, where the
diffractively produced state $X$ contains a gauge boson decaying to
a lepton pair.}
\label{fig:3-regge}      
\end{figure*}

\begin{figure*}[h]
\vspace*{-5mm}
\centering
 \includegraphics[width=7cm,clip]{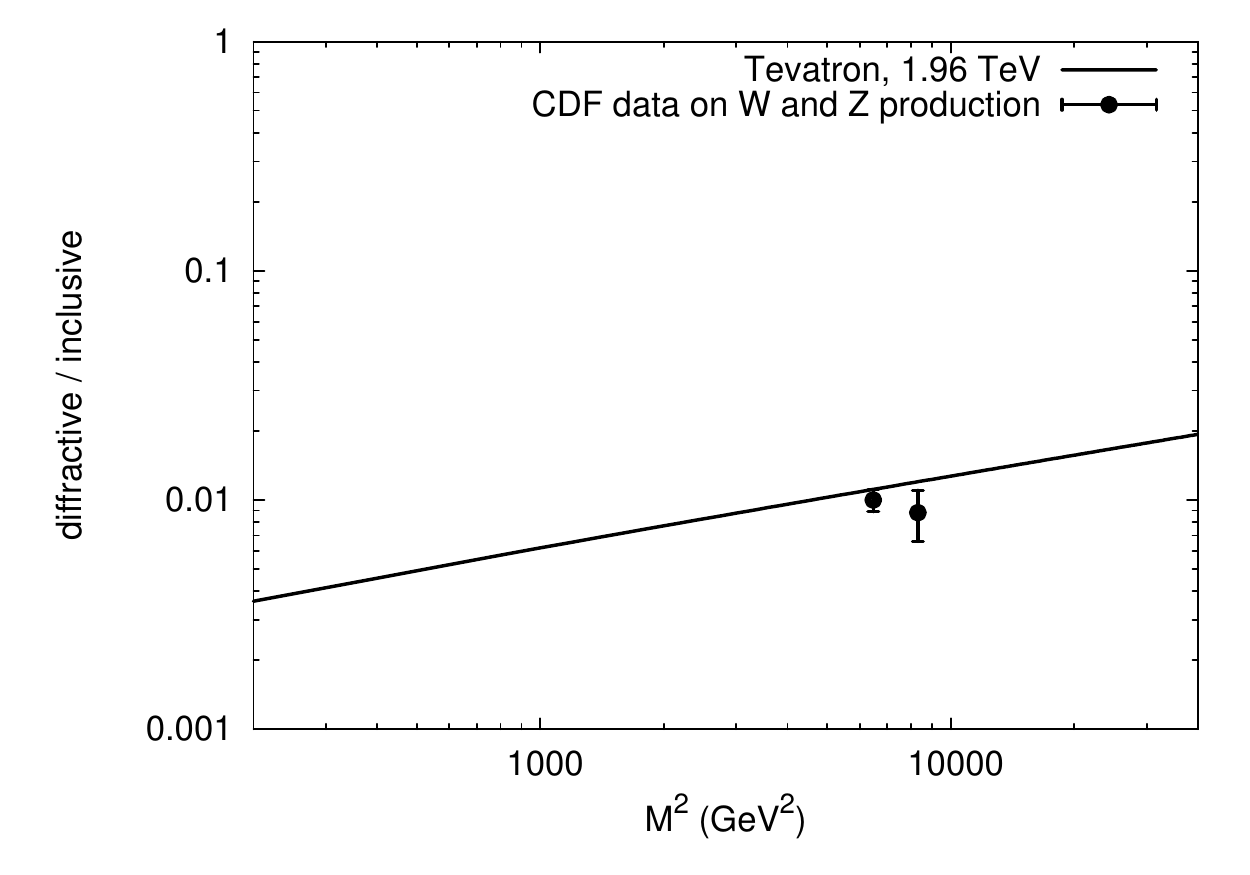}
\hspace*{1cm}
\includegraphics[width=4cm,clip]{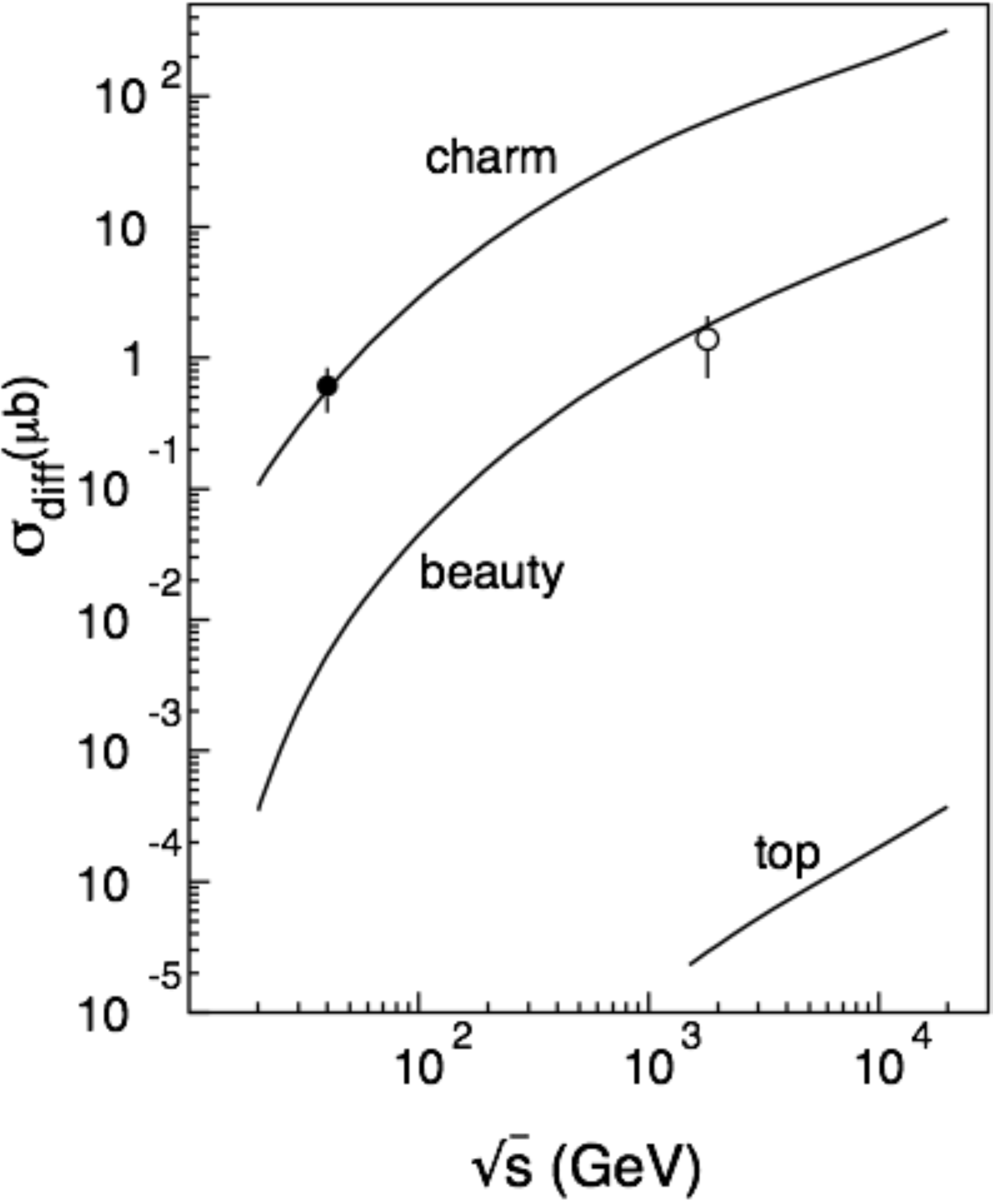}
\vspace*{-3mm}     
\caption{{\sl Left:} The diffractive-to-inclusive ratio vs dilepton invariant mass squared in comparison with the CDF measurements. {\sl Right:} Cross section of diffractive production of heavy flavors in comparison with the CDF data for charm and beauty (see details in \cite{hf-diff}).}
\label{fig:diff-data}      
\end{figure*}

The single diffractive process $pp\to Xp$ at large Feynman $x_F\to1$ of the recoil proton is described by the triple Regge graphs,
as is illustrated in Fig.~\ref{fig:3-regge}.
The results for the fractional diffractive cross sections of $Z$ and $W$ production is compared with the CDF measurements in Fig.~\ref{fig:diff-data} (left panel).

\section{Diffractive heavy flavor production}
\label{hf}

\begin{figure*}[h]
\centering
 \includegraphics[width=5cm,clip]{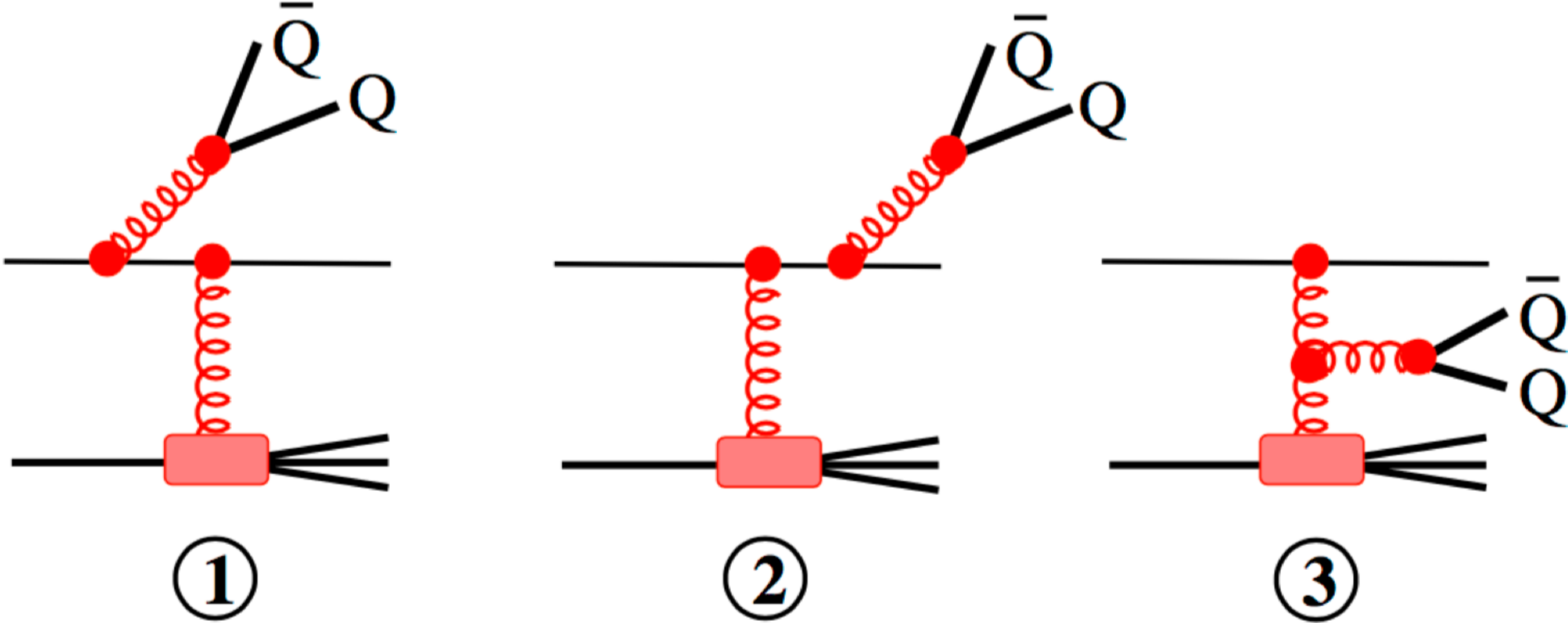}
 \hspace*{5mm}
 \includegraphics[width=3.5cm,clip]{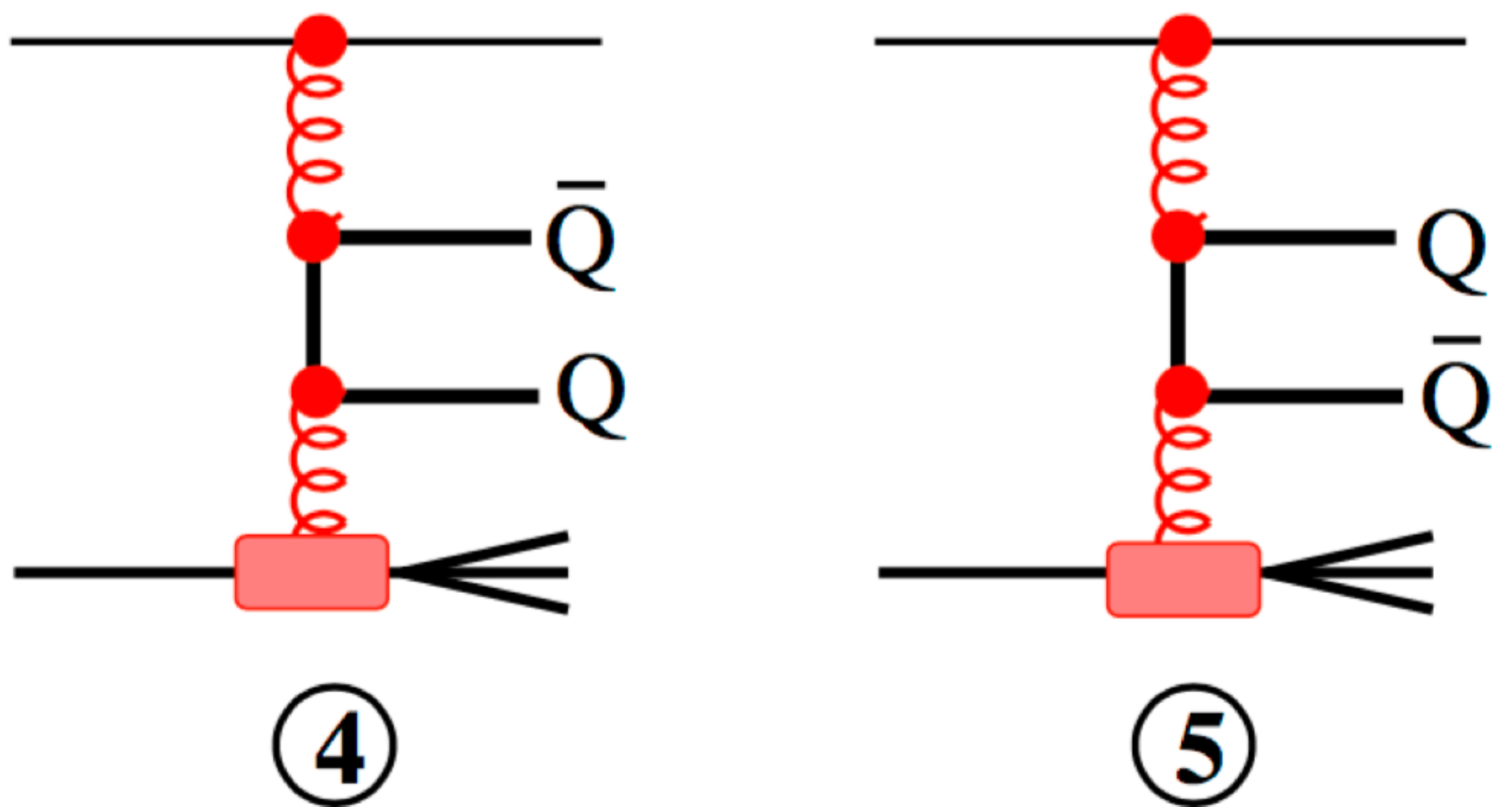}
\caption{Feynman graphs contributing to inclusive production
of a heavy quark pair.}
\label{fig:graphs}      
\end{figure*}

The dynamics of inclusive heavy flavor production can be classified as : (i) bremsstrahlung (like in DY); and
(ii) production mechanisms \cite{hf-diff}, in accordance with the Feynman graphs presented in Fig.~\ref{fig:graphs}. 
The amplitudes for the two mechanisms are expressed via the amplitudes $A_i$ corresponding to the graph numbering  in Fig.~\ref{fig:graphs}. The bremsstrahlung and production amplitudes have the form,
\beqn
A_{Br}&=&A_1+A_2+\frac{Q^2}{M^2+Q^2}\,A_3;
\label{120}
\\
A_{Pr}&=&\frac{M^2}{M^2+Q^2}\,A_3+A_4+A_5.
\label{140}
\eeqn

\begin{figure*}[h]
\centering
 \includegraphics[width=3.5cm,clip]{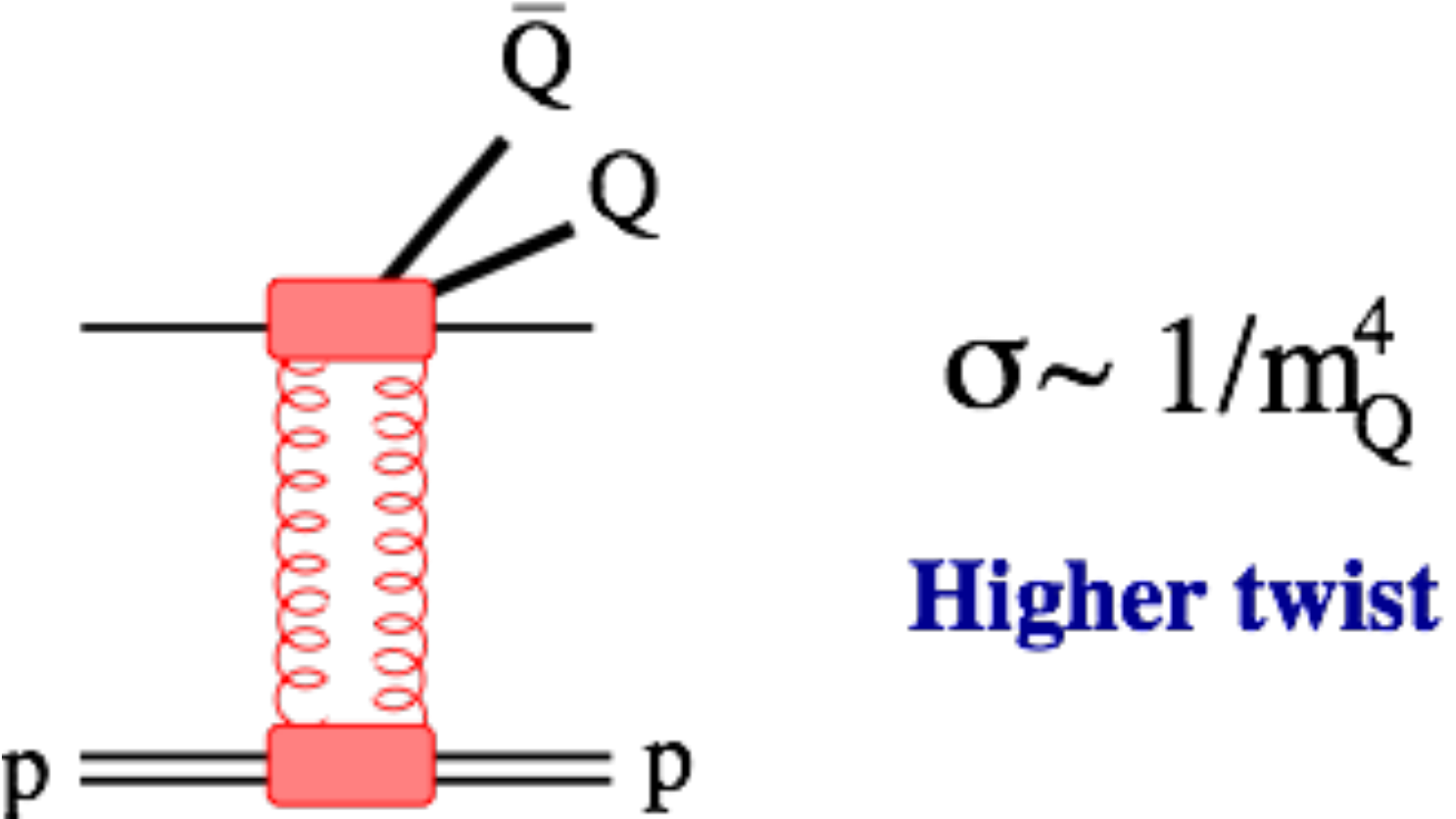}
 \hspace*{1cm}
 \includegraphics[width=5.5cm,clip]{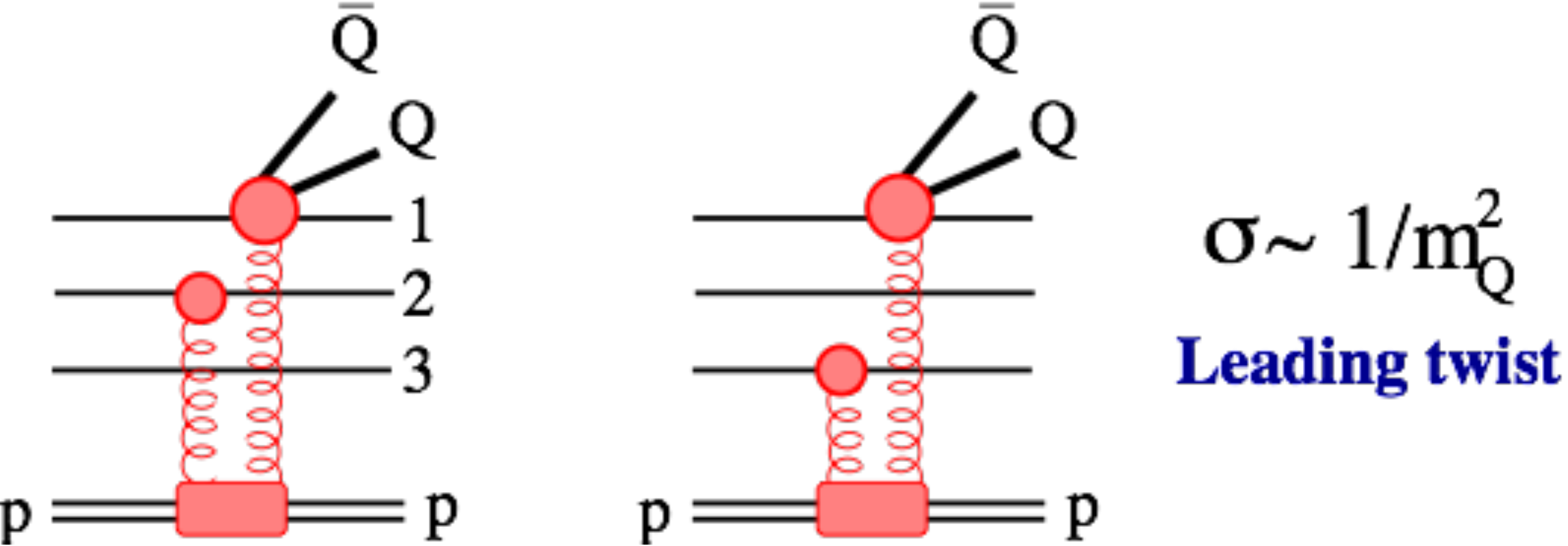}
\caption{{\sl Left:} Diffractive production of a heavy quark pair in a quark-proton collision.
{\sl Right:} Diffractive production of heavy quarks in a proton-proton collision.}
\label{fig:twists}      
\end{figure*}

For diffractive production one has to provide a colorless two-gluon exchange.
Diffractive excitation of a quark turns out to be a higher twist effect, as is depicted in Fig.~\ref{fig:twists}, left, similar to diffractive Drell-Yan.
In this case interaction with spectators again plays crucial role, providing a leading twist contribution, as is shown in Fig.~\ref{fig:twists}, right \cite{hf-diff}.
Numerically, the leading twist production mechanism is much larger 
compared with the bremsstrahlung mechanism.

The leading twist behavior $1/m_Q^2$
of the diffractive cross section is 
confirmed by CDF data, as is demonstrated in Fig.~\ref{fig:diff-data}, right panel.

\section{Diffractive Higgsstrahlung}
\label{higgs}

\begin{figure*}[h]
\centering
 \includegraphics[width=6cm,clip]{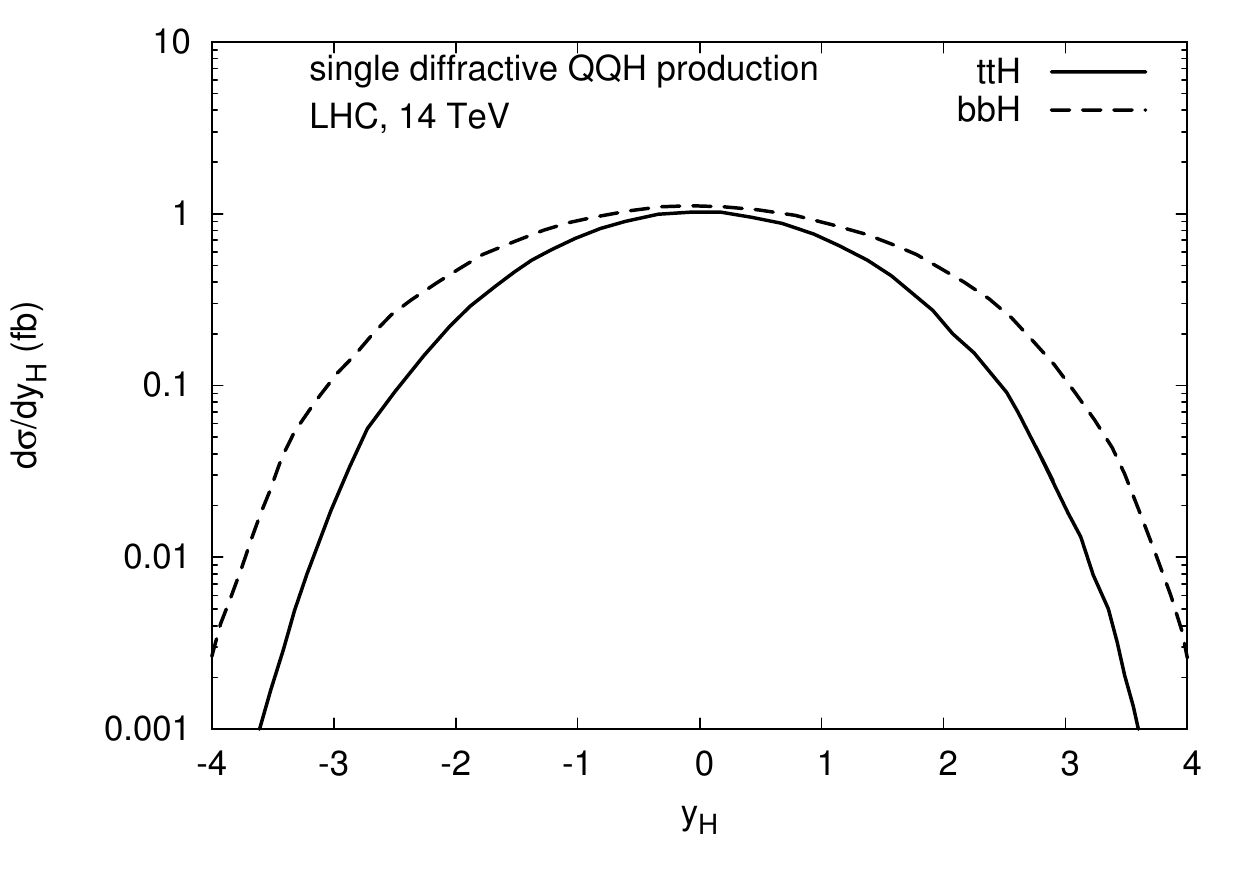}
 \includegraphics[width=5.5cm,clip]{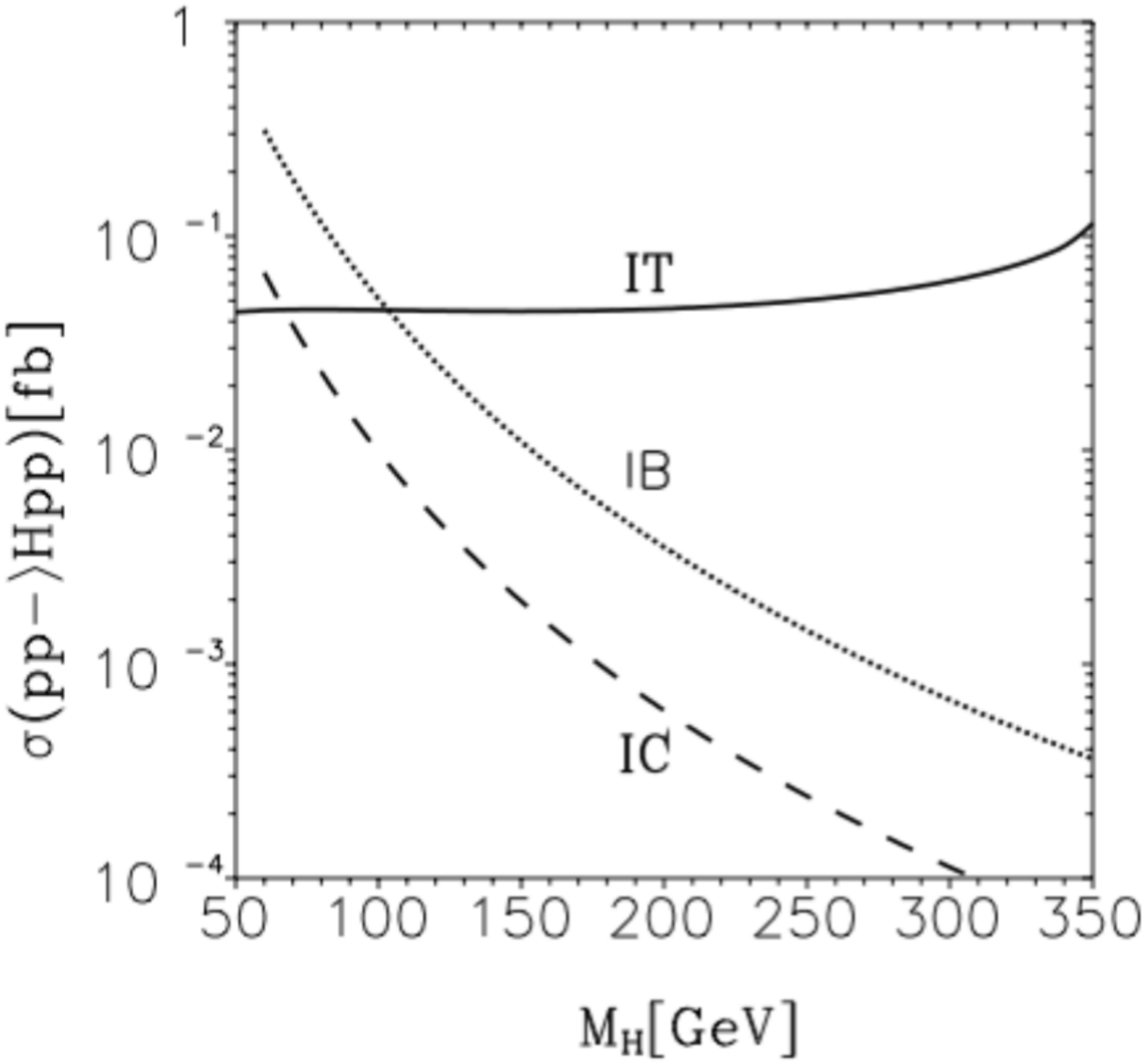}
\caption{{\sl Left:} The differential cross section of single diffractive Higgs boson production in association with heavy quark pair vs Higgs  rapidity (see details in \cite{higgs-r}).
{\sl Right:} The cross section of diffractive exclusive Higgs production, vs its mass, for different intrinsic heavy flavors \cite{brod1}.   }
\label{fig:higgs}      
\end{figure*}

Diffractive Higgsstrahlung is similar to diffractive DY. Z, W, since in all cases the radiated particle does not participate in the interaction. However, the Higgs decouples from light quarks, due to smallness of the coupling, so the cross section of higgsstrahlung by light hadrons is small.
Although light quark do not radiate Higgs directly, they can
do it via production of heavy flavors.
Therefore the mechanism is closely related to  
non-abelian diffractive quark production, presented in the previous section.
The rapidity dependent cross section of diffractive Higgs production, evaluated in \cite{higgs-r}, is plotted in Fig.~\ref{fig:higgs}, left.
The cross section is rather small, below $1$fb.

Higgs boson can also be diffractively produced  due to intrinsic heavy flavors in light hadrons. Exclusive Higgs production, $pp\to Hpp$, via coalescence of heavy quarks was evaluated in 
\cite{brod1,brod2}. 

The cross section of Higgs production was calculated fixing 1\% of intrinsic charm in the proton, and assuming that heavier flavors scale as $1/m_Q^2$ \cite{polyakov}. At the Higgs mass $125\GeV$ intrinsic bottom and top give comparable contributions,
as is demonstrated in Fig.~\ref{fig:higgs}, right.

\section{Summary}

Factorization of short and long-distance interactions is heavily broken in hard diffractive hadronic collisions.
In particular, forward diffractive radiation of direct photons, Drell-Yan dileptons, and gauge bosons Z, W, by a parton is forbidden. Nevertheless, a hadron can diffractively radiate in the forward direction due to a possibility of soft interaction with the spectators. This property of abelian radiation breaks down diffractive factorization resulting in a leading twist dependence on the boson mass,  $1/M^2$.

Non-abelian forward diffractive radiation of heavy flavors is permitted even for an isolated parton. However, interaction with spectators provides the  dominant contribution to the cross section. 
It comes from the interplay between large and small distances. Data well confirm the leading twist behavior.

Diffractive higgsstrahlung is possible due to a double-step process, via heavy quark production. Therefore, the main contribution comes for Higgs production in association with a heavy quark pair.
Another important contribution to  diffractive Higgs production comes from 
coalescence of intrinsic heavy quarks in the proton. For $M_H=125\GeV$ dominance of intrinsic bottom and top is expected.

\vspace{5mm}

{\bf Acknowledgement:}
This work was supported in part
by Fondecyt (Chile) grant 1130543 and Conicyt (Chile) grant Anillo ACT 1406.

\end{document}